# Absorption free superluminal light propagation in a three level pump-probe system

M. Mahmoudi [(1)], S. Worya Rabiei [(1)], L. Ebrahimi Zohravi [(1)] and M. Sahrai [(2)]

[(1)] Department of Physics, Zanjan University, P.O.Box 45195-313, Zanjan, Iran

[(2)] Research Institute for Applied Physics and Astronomy,

University of Tabriz, Tabriz, Iran

**Abstract**

We investigate the dispersion and the absorption properties of a weak probe field in a three-level pump-probe atomic system. It is shown that the slope of dispersion changes from positive to negative just with the intensity of the coherent or indirect incoherent pumping fields. It is demonstrated that the absorption free superluminal light propagation is appeared in this system.

Propagation of electromagnetic pulse in a dispersive medium has been a topic of recent study in the field of quantum coherence and interference [1-4]. Lord Rayleigh discussed the phenomenon of anomalous dispersion of a wave group in 1899 [5]. It is well known that the group velocity of a light pulse can be slowed down, or it can become greater than $c$ (the speed of light in vacuum) or even become negative in a transparent medium [6]. The effect of superluminality is that the emerging pulse has essentially the same shape and width as that of the incident wave packet, but its peak travels with a velocity higher than $c$ and even exits the medium before the incident pulse enters. This processes can be understood in terms of superposition and interference of traveling plane waves that have a distribution of frequencies and add up to form a narrow-band light pulse [7,8]. A rather simple mathematical proof shows that fast light behavior is completely consistent with Maxwell's equations describing pulse propagation through a dispersive material and hence does not violate Einstein's special theory of relativity (the special theory of relativity is based on Maxwell's equations) [9]. Intriguing question arise whether one can have a controlling parameter in a single experiment for switching from the normal to anomalous dispersion. Various schemes have been proposed to switching from subluminal to superluminal light propagation in an atomic medium. The most important key to successful experiments on subluminal and superluminal light propagation lies in its ability to control the optical properties of a medium by coherent or incoherent fields. The intensity control of group velocity has been attempted by using a lower level-coupling field in the three-level atomic system [10]. In another study, a scheme based on four-level electromagnetically induced transparency for switching from subluminal to superluminal light propagation via relative phase between driving fields was introduced [11]. Korsunsky et al. presented a theory of a continuous wave light propagation in a medium of atoms with a double $\Lambda$ configuration [12]. They have shown that, when the so-called multi-photon resonance condition is fulfilled, both absorptive and dispersive properties of such medium depend on the relative phase between the driving fields. In another study, the light propagation through closed-loop atomic medium beyond the multi-photon resonance condition has also been studied [13]. It is shown that the medium response oscillating in phase with the probe field, in general, is not phase-dependent and for parameters violating the multiphoton resonance condition, inducing a closed loop interaction contour is not advantageous.

It has also been demonstrated that the incoherent pump field has a major role in the controlling of the group velocity of the light pulse in a dispersive medium. The effect of the incoherent pumping field as well as spontaneously generated coherence (SGC) [14] on controlling the group velocity of probe field has been discussed [15, 16].

In view of many proposals, we note that the superluminality is accompanied by a considerable absorption. The proposal by Agarwal et al. [10] is notable for intensity control over the group velocity where the superluminality is accompanied by gain as in the experimental observation by Wang et al. [17] and the theoretical proposal by Wilson-Gordon et al. [18]. It has also been attempted to reduce the absorption as pulse propagates through the medium [11, 19].

In this article, employing an indirect incoherent pumping field, we show that the slope of dispersion can change from positive to negative just with the intensity of the coherent or incoherent pump fields. To illustrate the advantage of this model, we note that the absorption free superluminal light propagation is appeared in this system. Note that in the presence of a direct incoherent pump field [15] the superluminal light propagation is accompanied with the absorption, while with indirect incoherent pumping field the absorption free superluminal light propagation is obtained. To investigate the attenuation of the probe field through the medium, the absorption properties of the probe field is also discussed.

Let us consider a closed three-level atomic system with two well separated lower-levels $|1\rangle$ and $|2\rangle$, and upper level $|3\rangle$ as shown in Fig. 1(a). A strong coherent coupling field of frequency $\nu_c$ with the Rabi-frequency $\Omega_c = \frac{E_c \wp_{23}}{2\hbar}$ drives the $|2\rangle \leftrightarrow |3\rangle$ transition, while a weak tunable probe field of frequency $\nu_p$ with Rabi-frequency $\Omega_p = \frac{E_p \wp_{13}}{2\hbar}$ applies to the $|1\rangle \leftrightarrow |3\rangle$ transition. Here $\wp_{ij}$ are the atomic dipole moments, while $E_c$, and $E_p$ are the amplitude of the coupling and the probe fields, respectively. An indirect incoherent pump field with the pumping rate $R$ applies to the $|1\rangle \leftrightarrow |3\rangle$ transition to generate a gain at the probe field transition frequency. The spontaneous decay rates from level $|3\rangle$ to the ground levels $|1\rangle$ and $|2\rangle$ are denoted by $\gamma_1$ and $\gamma_2$, respectively.

The density matrix equations of motion in the rotating wave approximation and in the rotating frame are

$$\dot{\rho}_{11} = \gamma_1 \rho_{33} + i(\Omega_p \rho_{31} - \Omega_p^* \rho_{13}) - R\rho_{11},$$

$$\dot{\rho}_{22} = \gamma_2 \rho_{33} + i(\Omega_c \rho_{32} - \Omega_p^* \rho_{23}),$$

$$\dot{\rho}_{21} = -[\frac{R}{2} + i(\Delta_c - \Delta_p)]\rho_{21} + i\Omega_c \rho_{31} - i\Omega_p^* \rho_{23},$$

$$\dot{\rho}_{31} = [i\Delta_p - \frac{1}{2}(\gamma_1 + \gamma_2 + R)]\rho_{31} + i\Omega_c^* \rho_{21} - i\Omega_p^*(\rho_{33} - \rho_{11}),$$

$$\dot{\rho}_{32} = [i\Delta_c - \frac{1}{2}(\gamma_1 + \gamma_2)]\rho_{32} + i\Omega_p^* \rho_{12} - i\Omega_c^*(\rho_{33} - \rho_{22}),$$

$$\rho_{11} + \rho_{22} + \rho_{33} = 1, \tag{1}$$

where the detuning of the pump and the probe fields are defined as $\Delta_c = \nu_c - \omega_{32}$, and $\Delta_p = \nu_p - \omega_{31}$. Note that the interference due to the different spontaneous emission channels, the spontaneously generated coherence, has been ignored [14].

In the following we discuss the response of the atomic system to the applied field by defining the susceptibility, $\chi$, as [20]

$$\chi = \frac{2N\wp_{13}}{\varepsilon_0 E_p} \rho_{31}. \tag{5}$$

Here N is the atom number density in medium, and $\chi = \chi' + i\chi''$. The real and imaginary parts of $\chi$ correspond to the dispersion and the absorption of the weak probe field, respectively. The group velocity of the weak probe field is then given by [10, 17]

$$v_g = \frac{c}{1 + 2\pi\chi'(v_p) + 2\pi v_p \frac{\partial \chi'(v_p)}{\partial v_p}} = \frac{c}{n_g}, \qquad (6)$$

where $c$ is the speed of light in the vacuum, $\chi'(v_p)$ is the real part of $\chi$, and $n_g$ shows the group index. For the realistic example we consider $\frac{2N\wp_{13}}{\varepsilon_0 E_p} \cong 1$ [15].

Before showing the numerical results of equations (1), we drive an analytical expression to obtain the necessary intensity of coupling field for switching the group velocity of probe field from subluminal to superluminal.

The steady state solutions for the weak probe field approximation, i.e. $\Omega_p \ll \gamma$, and for $\Delta_c = 0$, is

$$\rho_{31} = \frac{4\Omega_c^2 \Omega_p [2\Delta_p(1-R) - iR^2]}{[R + 2\Omega_c^2(1+2R)][4\Omega_c^2 + R^2 + 2R(1-2i\Delta_p) - 4\Delta_p(i+\Delta_p)]}, \qquad (2)$$

where $\gamma_1 = \gamma_2 = \gamma$ and all parameters are reduced to dimensionless units through scaling by $\gamma$. For the small probe field detuning, i.e. $\Delta_p \ll \gamma$, the real part of $\rho_{31}$ is given by:

$$\text{Re}[\rho_{31}] = \frac{8\Omega_c^2 \Omega_p [R^3 + R^2 + 2R + 4\Omega_c^2(1-R)]}{[R + 2\Omega_c^2(1+2R)](4\Omega_c^2 + R^2 + 2R)^2} \Delta_p. \qquad (3)$$

An investigation of Eq. (3) shows that the slope of dispersion depends on the intensity of the coherent and indirect incoherent pumping fields. For the weak indirect incoherent pumping rate, $R < \gamma$, the slope of dispersion around zero probe field detuning is definitely positive that is correspond to the subluminal light propagation. For $R > \gamma$, the slope of dispersion can change from positive to negative just by increasing the intensity of coupling field.

For a given $R$, the necessary value of $\Omega_c$ to change the slope of dispersion from positive to negative, is obtain by

$$(\Omega_c)_{necessary} = \frac{1}{2}\sqrt{\frac{R^3 + R^2 + 2R}{R-1}}. \tag{4}$$

For $\Omega_c < (\Omega_c)_{necessary}$, the slope of dispersion around zero detuning is positive, while for $\Omega_c > (\Omega_c)_{necessary}$ it becomes negative.

In Fig. 2, we display the subluminal and superluminal regions via the Rabi frequency of coupling field and indirect incoherent pumping rate (see eq. (4)). The superluminal region is shown with dark color and has a minimum point at $(\Omega_c)_{min} = 1.99\gamma$. This is the required minimum intensity of the coupling field to switching the group velocity from subluminal to superluminal light propagation. In the case of $\Omega_c < (\Omega_c)_{min}$ (e.g. dashed line), the normal dispersion can not switch to anomalous dispersion even by changing of the indirect incoherent pumping field rate. So, the superluminal light propagation is impossible at this region. For $\Omega_c > (\Omega_c)_{min}$ (e.g. dash-dotted line), the superluminal region is limited by two incoherent pumping rates $R_1$ and $R_2$ ($R_1$ and $R_2$ are the roots of equation (4)). For $R_1 < R < R_2$ the slope of the dispersion around zero probe field detuning is negative, while for $R < R_1$ or $R > R_2$ it becomes positive.

We now summarize our results for the steady state behavior of the system by using equations (1) and (5). We assume $\gamma_1 = \gamma_2 = \gamma$, and all figures are plotted in the unit of $\gamma$. For simplicity, the Rabi-frequencies are considered to be real. In our notation if $\text{Im}[\chi] < 0$, the system exhibits gain for the probe field, while for $\text{Im}[\chi] > 0$, the probe field is attenuated. First, we investigate the effect of coherent coupling field on the dispersion and the absorption properties of the weak probe field. In Fig. 3, we display the dispersion (a) and the absorption (b) properties of the probe field versus probe field detuning for various intensity of coupling field. The common parameters are $\Omega_p = 0.01\gamma$, $\Delta_c = 0.0$, $R = 1.5\gamma$. Fig. 3(a) shows that for $\Omega_c < (\Omega_c)_{necessary}$ (Solid), i.e. $\Omega_c = 1.25\gamma$, the slope of dispersion is positive corresponding to the subluminal light propagation. For $\Omega_c = (\Omega_c)_{necessary} = 2.08\gamma$ the slope of dispersion become zero, while for $\Omega_c > (\Omega_c)_{necessary}$ the slope of dispersion changes to the negative. The effect of the Rabi-frequency of coupling field on the absorption is shown in Fig .3(b). It is realized that for all values of $\Omega_c$, the system shows a gain profile and with increasing the intensity of coupling field the initial gain profile separates into two dips.

Now, we investigate the effect of indirect incoherent pumping field on the pulse propagation.

In Fig. 4, we show the dispersion (a) and the absorption (b) properties of the probe field versus probe detuning for different indirect incoherent pumping rates. For $R = 0.8\gamma$ (solid lines) the slope of dispersion around zero probe field detuning is positive (Fig.3 (a)). With increasing the indirect incoherent pumping rate, i.e. $R = 1.14\gamma$ (dashed lines), the slope of dispersion is zero and for $R = 1.5\gamma$ (dash-dotted lines) it becomes negative. Fig. 3(b) shows that the subluminal and superluminal light propagation around zero probe detuning is accompanied by a considerable gain. Thus, the indirect incoherent pumping rate has a major role in switching of the light propagation from subluminal to superluminal.

In the following we present physical mechanism of the above results by defining the dressed states. In the absence of indirect incoherent pumping field, i.e. R=0, the atom-field states are

$$|\pm\rangle = \frac{1}{\sqrt{2}}(|3\rangle \pm |2\rangle), \tag{7}$$

with eigenvalues

$$\lambda_\pm = \mp \hbar \Omega_c. \tag{8}$$

The atom-field states in the new basis are shown in Fig.1 (b). It has been shown that the two absorption electromagnetically induced transparency (EIT) peaks should be appeared in the absorption spectrum [21]. Note that for R=0, the relation (2) converts to

$$\rho_{31} = \frac{\Omega_p \Delta_p}{\Omega_c^2 - \Delta_p^2 - i\Delta_p}, \tag{9}$$

which describes the electromagnetically induced transparency in a simple three-level pump-probe system [1, 21].

By employing the indirect incoherent pumping field to $|1\rangle - |3\rangle$ transition, the dressed states populate, and the probe field absorption switches to the gain. For the weak coupling field the two closely dressed states establish a gain dip in the spectrum (solid line in Fig.3 (b)), but with increasing the intensity of coupling field the gain profile start to split into two dips (see dashed and dash-dotted lines in Fig. 3 (b) and Fig.4 (b)).

In Fig. 5, we show the effect of coherent (a) and incoherent pumping fields (b) on the group index. The common selected parameters are $\Omega_p = 0.01\gamma$, $\Delta_c = 0.0$, $\Delta_p = 0.0$. Fig. 5 (a) shows that for the weak indirect incoherent pumping field, i.e., $R = \gamma$ (solid), the group index is positive, while for the larger indirect incoherent pumping rate, i.e., $R = 1.14\gamma$ (dashed) and $R = 1.5\gamma$ (dash-dotted), the group index become negative as the intensity of coupling field increases. In point (I) of the dash-dotted curve the group index becomes unit that corresponds to the dashed line of Fig. 3. The point (II) of the dash-dotted curve is corresponding to dash-dotted lines of Fig. 3 and Fig. 4. In the point (III) of dashed line the group index is also unit that is correspond to the dashed line of fig.4.

Fig. 5(b) displays the group index versus indirect incoherent pumping rate. For $\Omega_c = 1.99\gamma$ (solid) the group index is positive, while for larger Rabi-frequencies, i.e., $\Omega_c = 2.08\gamma$ (dashed) and $\Omega_c = 3\gamma$ (dash-dotted), it changes from positive to negative and back to positive as the indirect incoherent pumping rate is increases. The group index will finally reach to the unit for the large indirect incoherent pumping field.

In conclusion, we investigate the effect of the intensity of coherent coupling and indirect incoherent pumping fields on the group velocity of a weak probe field. It is shown that the slope of dispersion not only depends on the intensity of coherent coupling field, but is dramatically depends on the intensity of the indirect incoherent pumping field as well. Moreover, it is demonstrated that the weak probe field does not attenuate as it passes through the medium, and the superluminal light propagation is accompanied by a gain.

**Figure captions**

Fig.1. Proposed level scheme. a) Three –level Λ-type system driven by the strong pump and weak probe fields. The dashed line shows the indirect incoherent pumping field. b) Equivalent level schemes in terms of the dressed state basis for $R = 0$.

Fig. 2. The subluminal and superluminal regions via the Rabi frequency of coupling field and indirect incoherent pumping rate (see Eq. (4)). The superluminal region is shown with dark colure.

Fig.3. Real (a) and imaginary (b) parts of susceptibility versus probe field detuning for the parameters $\gamma_1 = \gamma_2 = \gamma$, $\Omega_p = 0.01\gamma$, $R = 1.5\gamma$, $\Omega_c = 1.25\gamma$ (solid), $2.08\gamma$ (dashed), $3.0\gamma$ (dash-dotted).

Fig.4. Real (a) and imaginary (b) parts of susceptibility versus probe field detuning for the parameters $\gamma_1 = \gamma_2 = \gamma$, $\Delta_c = 0$, $\Omega_p = 0.01\gamma$, $\Omega_c = 3.0\gamma$, $R = 0.8\gamma$ (solid), $1.14\gamma$ (dashed), $1.5\gamma$ (dash-dotted).

Fig.5. Group index ($\frac{c}{v_g} - 1$) versus coherent coupling Rabi frequency (a) and indirect incoherent pumping rate (b) for the parameters $\gamma_1 = \gamma_2 = \gamma$, $\Delta_c = 0$, $\Omega_p = 0.01\gamma$, a) $R = \gamma$ (solid), $1.14\gamma$ (dashed), $1.5\gamma$ (dash-dotted), b) $\Omega_c = 1.99\gamma$ (solid), $2.08\gamma$ (dashed), $3.0\gamma$ (dash-dotted).

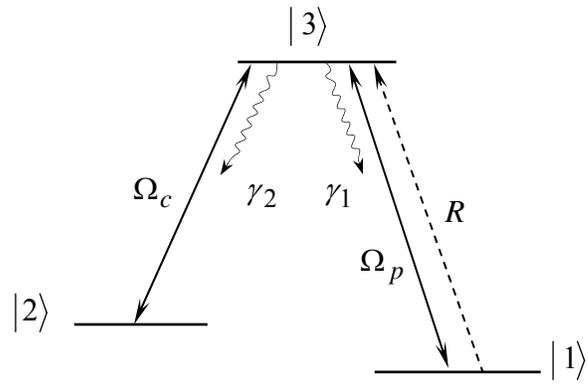

Fig. 1 (a)

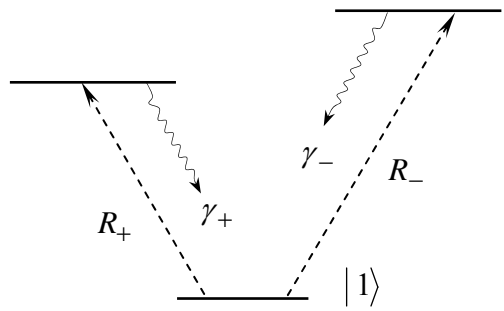

Fig. 1 (b)

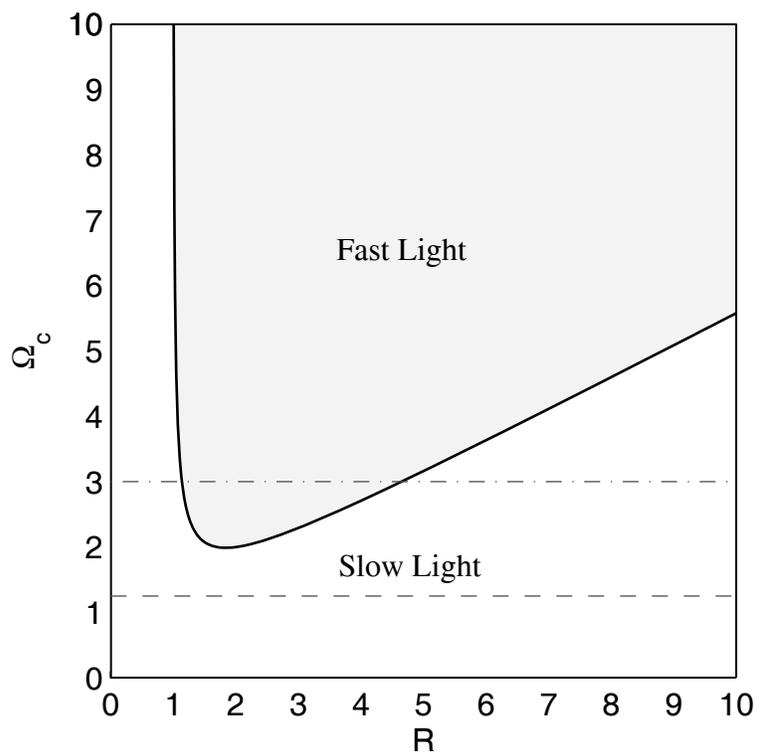

Fig. 2

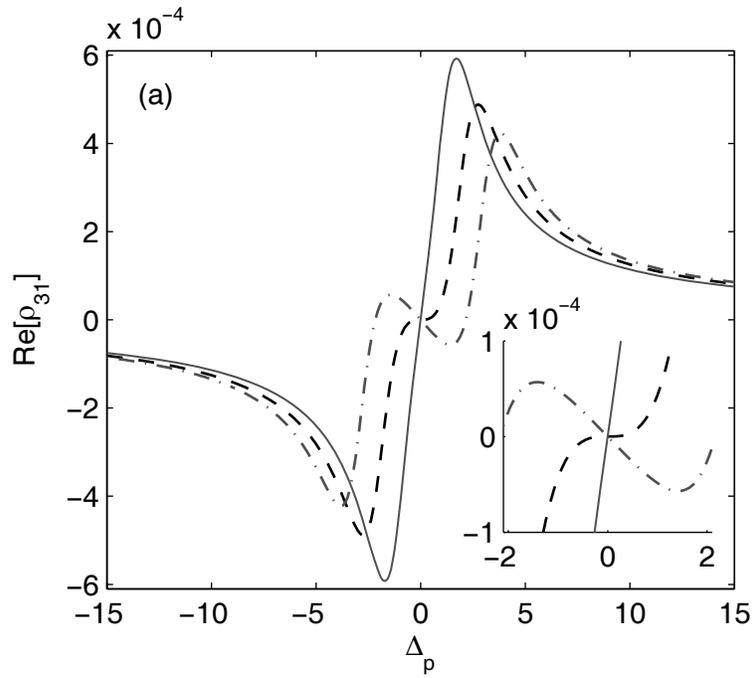

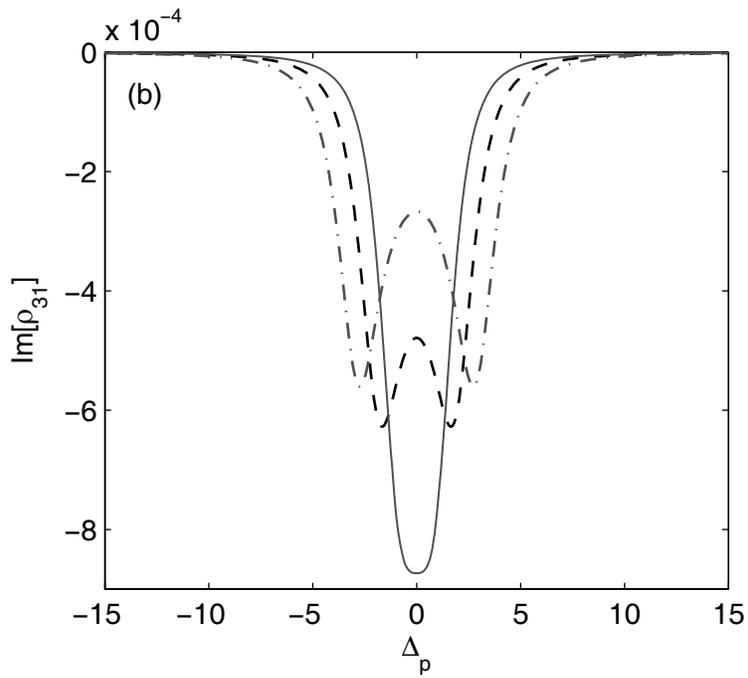

Fig. 3

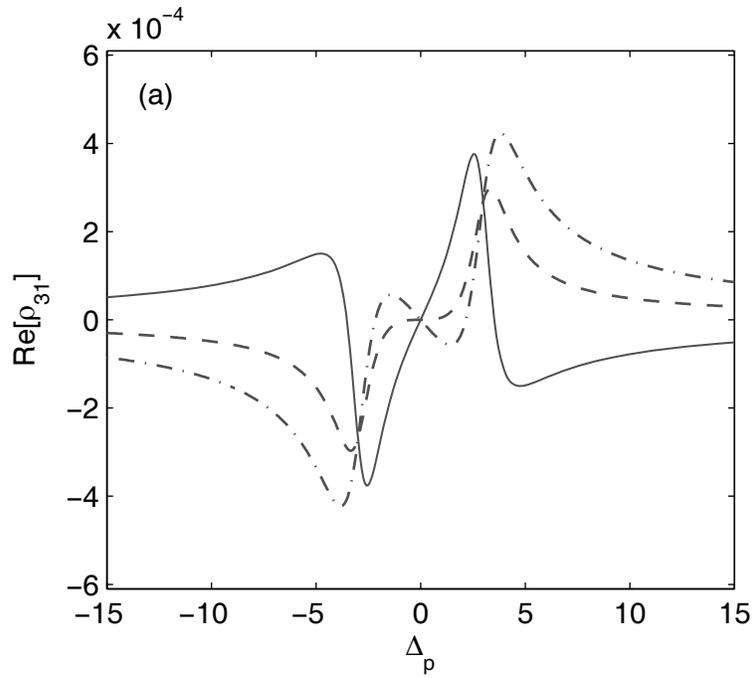

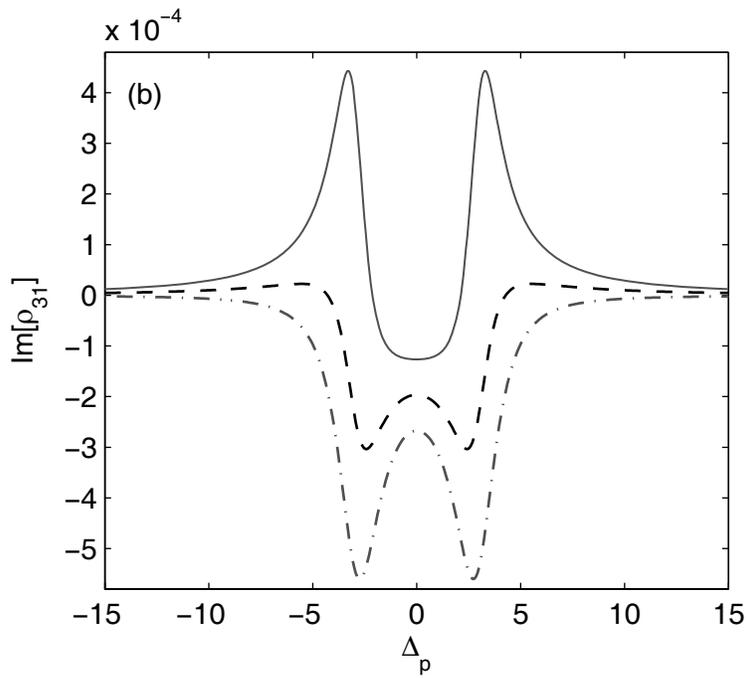

Fig.4

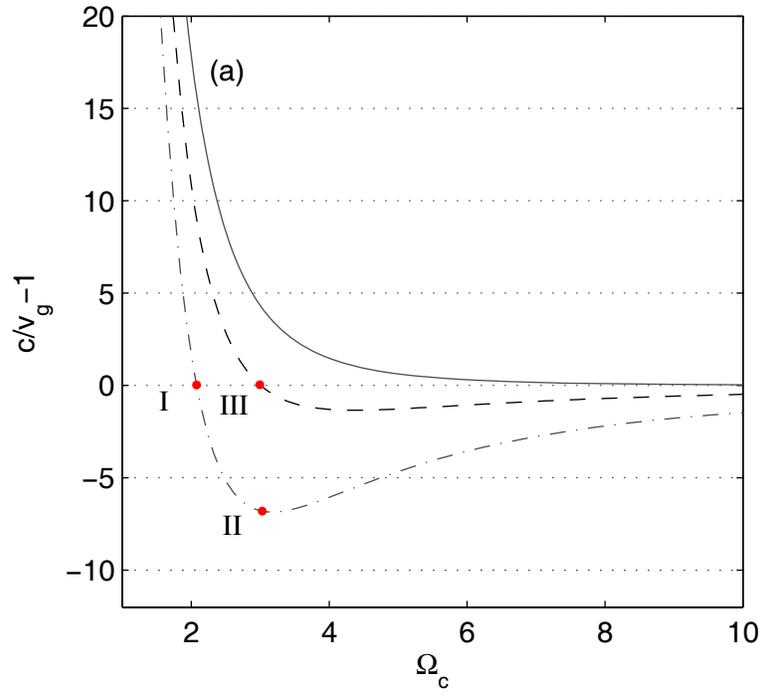

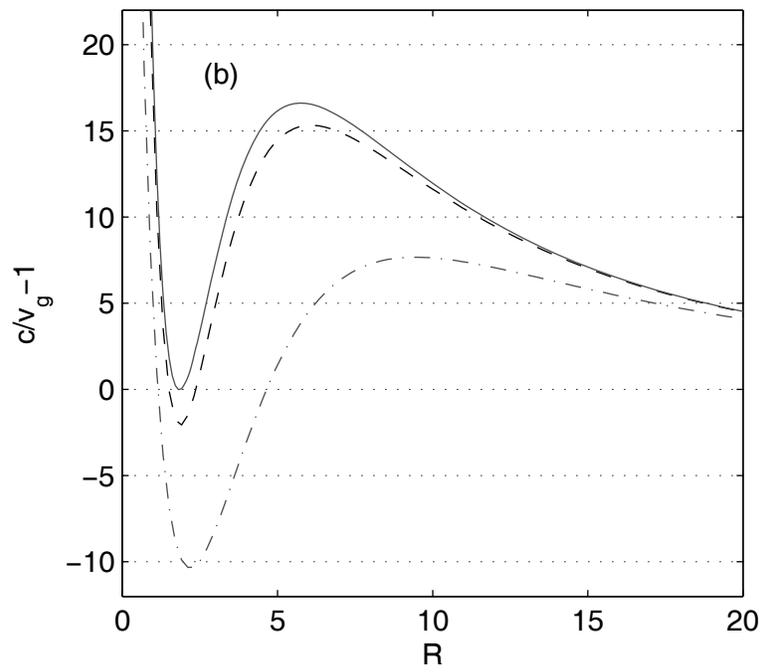

Fig. 5